\documentclass[prd,twocolumn,superscriptaddress,showpacs,nofootinbib,preprintnumbers]{revtex4}
\usepackage{amsmath}
\usepackage{amsfonts}
\usepackage{graphicx}
\usepackage{dcolumn}
\usepackage{hyperref}

\def\be{\begin{equation}}
\def\ee{\end{equation}}
\def\bea{\begin{eqnarray}}
\def\eea{\end{eqnarray}}

\def\5{\overline 5}

\begin{document}
\title{The fate of (phantom) dark energy universe with string curvature 
corrections}
\author{M.~Sami}
\altaffiliation[On leave from:]{ Department of Physics, Jamia Millia, New Delhi-110025} 
\affiliation{IUCAA, Post Bag 4, Ganeshkhind,
Pune 411 007, India}
\author{Alexey Toporensky}
\affiliation{Sternberg Astronomical Institute, Moscow State University, Universitetsky 
Prospekt, 13, Moscow 119899, Russia}
\author{Peter V. Tretjakov}
\affiliation{Sternberg Astronomical Institute, Moscow State University, Universitetsky 
Prospekt, 13, Moscow 119899, Russia}
\author{Shinji Tsujikawa}
\affiliation{Department of Physics, Gunma National College of
Technology, Gunma 371-8530, Japan}

\begin{abstract}
    
We study the evolution of (phantom) dark energy universe by taking into 
account the higher-order string corrections
to Einstein-Hilbert action with  fixed dilaton and modulus fields.
While the presence of a cosmological constant
gives stable de-Sitter fixed points in the cases of 
heterotic and bosonic strings,
no stable de-Sitter solutions exist when a phantom fluid is present.
We find that the universe can exhibit a Big 
Crunch singularity with a finite time for type II string, 
whereas it reaches a Big Rip singularity for heterotic and 
bosonic strings. Thus the fate of dark energy universe 
crucially depends upon the type of string theory under consideration.

\end{abstract}


\maketitle
\vskip 1pc

\section{Introduction}

Recent observations suggest that the current universe is 
dominated by dark energy responsible for 
an accelerated expansion \cite{obser}. 
The equation of state parameter $w$
for dark energy lies in a narrow region around $w=-1$
and may even be smaller than $-1$ \cite{Caldwell}.
When $w$ is less than $-1$, dubbed as phantom dark 
energy, the universe ends up with a Big Rip 
singularity \cite{Sta,CKW} which is characterized by the divergence 
of curvature of the universe after a finite interval of time
(see Refs.~\cite{phantom0,phantom}).

The energy scale may grow up to the Planck scale in the 
presence of phantom dark energy.
This means that higher-order 
curvature or quantum corrections can be important around 
the Big Rip. For example, quantum corrections coming from 
conformal anomaly are taken into account in Refs.~\cite{Nojiri}
for dark energy dynamics. It was found that
such corrections can moderate the singularity 
by providing a negative energy density \cite{NOT}.
Thus it is important to implement quantum effects in order 
to predict the final fate of the universe.

In low-energy effective string theory there exist higher-curvature corrections
to the usual scalar curvature term.
The leading quadratic correction corresponds to the product of 
dilaton/modulus and Gauss-Bonnet (GB)
curvature invariant \cite{BD}.
The GB term is topologically invariant in four dimensions and 
hence does not contribute
to dynamical equations of motion 
if the dilaton/modulus field is constant \cite{BB}.
Meanwhile it affects the cosmological dynamics 
in presence of dynamically evolving dilaton and modulus fields.
The possible effects of the GB term for early universe cosmology 
and  black hole physics were
investigated in Refs. \cite{GBearly,AP}.
Lately the GB correction was applied to the study of cosmological 
dynamics of dark energy \cite{NOS}.

When the dilaton and mudulus are fixed, it is 
important to implement third and next-order string 
curvature corrections \cite{BB}. 
This can change the resulting cosmological 
dynamics drastically as it happens in the context of inflation 
\cite{Ohta} and
black holes \cite{PKA}.
The goal of the present paper is to study the effect of next-to-leading order
string corrections to the cosmological dynamics around the Big Rip
singularity with an assumption that the dilaton and the modulus are stabilized.
We would also investigate the existence and the stability
of de-Sitter solutions in the presence of a cosmological constant.
We shall consider three types of string corrections
and study the fate of the universe accordingly.

\section{Evolution equations}

Let us consider the Einstein-Hilbert action in low-energy effective 
string theory:
\begin{eqnarray}
\label{action}
{\cal S} = \int d^D x \sqrt{-g}\left[R+
{\cal L}_{c}+\ldots\right]\,,
\end{eqnarray}
where $R$ is the scalar curvature and  ${\cal L}_{c}$ is
the string correction which is given by \cite{BB}
\begin{eqnarray}
{\cal L}_{c}=c_1\alpha^\prime e^{- 2 \phi}
{\cal L}_2 + c_2 \alpha^{\prime 2} e^{- 4 \phi}{\cal L}_3 + c_3
\alpha^{\prime 3} e^{- 6 \phi}{\cal L}_4 \,,
\end{eqnarray}
where $\alpha'$ is the string expansion parameter, $\phi$ is the 
dilaton field, and
\begin{eqnarray} 
& & {\cal L}_2 =\Omega_2\,, \\
& & {\cal L}_3 = 2 \Omega_3 +
R^{\mu\nu}_{\alpha \beta}
R^{\alpha\beta}_{\lambda\rho}
R^{\lambda\rho}_{\mu\nu}\,, \\
& & {\cal L}_4= {\cal L}_{41} -\delta_{H} {\cal L}_{42}
-\frac{\delta_{B}}{2}{\cal L}_{43}\,,
\end{eqnarray}
with
\begin{eqnarray} 
\hspace*{-1.0em}\Omega_2 &=& R^2-4R_{\mu \nu}R^{\mu \nu}+
R_{\mu \nu \alpha \beta}R^{\mu \nu \alpha \beta}, \\
\hspace*{-1.0em} \Omega_3 &=& R^{\mu\nu}_{\alpha \beta}
R^{\alpha\beta}_{\lambda\rho}R^{\lambda\rho}_{\mu\nu}
-2R^{\mu\nu}_{\alpha\beta}R_\nu^{\lambda\beta\rho}
R^\alpha_{\rho\mu\lambda} \nonumber \\
& & +{3\over 4} R R_{\mu\nu\alpha\beta}^2 + 
6  R^{\mu\nu\alpha\beta}R_{\alpha\mu}R_{\beta\nu} \nonumber \\
& &+ 4R^{\mu\nu}R_{\nu\alpha}R^\alpha_{\phantom{\alpha}\mu} - 6
R R_{\alpha\beta}^2 + \frac{R^3}{4},\\
\hspace*{-1.0em}  {\cal L}_{41} &=& \zeta(3)
R_{\mu\nu\rho\sigma}R^{\alpha\nu\rho\beta}\left(
R^{\mu\gamma}_{\delta\beta}
R_{\alpha\gamma}^{\delta\sigma}
- 2  R^{\mu\gamma}_{\delta\alpha}
R_{\beta\gamma}^{\delta\sigma} \right), \\
\hspace*{-1.0em} {\cal L}_{42} &=& {1\over 8} \left(
R_{\mu\nu\alpha\beta} R^{\mu\nu\alpha\beta}\right)^2
 +{1\over 4}  R_{\mu\nu}^{\gamma\delta}
R_{\gamma\delta}^{\rho\sigma}
R_{\rho\sigma}^{\alpha\beta}
R_{\alpha\beta}^{\mu\nu} \nonumber \\
& &- {1\over 2} R_{\mu\nu}^{\alpha\beta}
R_{\alpha\beta}^{\rho\sigma}
R^\mu_{\sigma\gamma\delta}
R_\rho^{\nu\gamma\delta}
- R_{\mu\nu}^{\alpha\beta}
R_{\alpha\beta}^{\rho\nu}
R_{\rho\sigma}^{\gamma\delta}
R_{\gamma\delta}^{\mu\sigma}, \\
\hspace*{-1.0em}  {\cal L}_{43} &=& \left(
R_{\mu\nu\alpha\beta}R^{\mu\nu\alpha\beta}\right)^2
- 10  R_{\mu\nu\alpha\beta}
R^{\mu\nu\alpha\sigma}
R_{\sigma\gamma\delta\rho}
R^{\beta\gamma\delta\rho} \nonumber \\
& &- R_{\mu\nu\alpha\beta}
R^{\mu\nu\rho}_{\sigma}
R^{\beta\sigma\gamma\delta}
R_{\delta\gamma\rho}^{\alpha}\,.            
\end{eqnarray}
Here one has $\delta_{H(B)}=1$ for heterotic (bosonic) string and zero otherwise. 
The Gauss-bonnet term, $\Omega_{2}$, as well as  the Euler density, $\Omega_{3}$, 
does not contribute to the background equation of motion for $D=4$ 
unless the dilaton is dynamically evolving.
The coefficients $(c_1,\ c_2,\ c_3)$  are different depending 
on string theories \cite{BB}. We have
$(c_{1}, c_{2}, c_{3})=(0, 0, 1/8), (1/8, 0, 1/8), (1/4, 1/48, 1/8)$
for type II, heterotic, and bosonic strings, respectively. 
In the case of type II string with $D=4$, for example,  
only the ${\cal L}_{41}$ term affects the dynamical evolution 
of the system.

We shall consider the flat Friedmann-Robertson-Walker metric
with a lapse function $N(t)$:  
\be
\label{metric}
ds^2=-N(t)^2dt^2+a(t)^2\sum_{i=1}^{d}
(dx^i)^2\,,
\ee
where $d=D-1$.
The Ricci tensors under this metric are given in the Appendix.
In what follows we shall consider the case of $D=4$ under the 
assumption that the modulus field which corresponds to the radius of 
extra dimensions is stabilized after the compactification to 
four dimensions. 
Then we find 
\begin{eqnarray}
\label{L3}
{\cal L}_3&=&\frac{24}{N^6}(H^6+I^3)-\frac{72\dot{N}}
{N^7}HI^2\,, \\
\label{L41}
{\cal L}_{41} &=& -\frac{6\zeta(3)}{N^8}
\left(3H^8+4H^4I^2+4H^2I^3+I^4\right) \nonumber \\
& & +\frac{6\zeta(3)\dot{N}}{N^9}
\left(8H^5I+12H^3I^2+4HI^3\right)\,, \\
\label{L42}
{\cal L}_{42} &=& -\frac{6}{N^8}
\left(5H^8+2H^4I^2+5I^4\right)  \nonumber \\
& &+\frac{6\dot{N}}{N^9}
\left(4H^5I+20HI^3\right)\,, \\
\label{L43}
{\cal L}_{43}&=& -\frac{6}{N^8} 
\left(60H^8+32H^4I^2+ 60I^4 \right)
\nonumber \\
& & +\frac{6 \dot{N}}{N^9} 
\left(64H^5I+240HI^3\right)\,,
\end{eqnarray}
where $H \equiv \dot{a}/a$ is the Hubble rate and 
$I$ is defined by $I \equiv H^2+\dot{H}$. 
It should be noted that, in the case of de-Sitter space time, the expressions 
(\ref{L3})-(\ref{L43}) reduce to their counterparts given by 
Eqs.~(\ref{dL1})-(\ref{dL3}) in the Appendix.

We shall implement the contribution of a barotropic perfect 
fluid to the action (\ref{action}).
The equation of state parameter, $w_m=p_m/\rho_m$,
is assumed to be constant.
Our main interest is to study the final fate of universe filled with
a phantom-type fluid ($w_{m}<-1$).
In this case the universe eventually reaches a 
Big Rip singularity \cite{Caldwell} with a divergent Hubble 
rate in the absence of higher-curvature terms. 
We are interested in the effect of string curvature corrections
to the cosmological evolution around the Big Rip.

Varying the action (\ref{action}) with respect to $N$, we find
\begin{eqnarray}
\label{Hami}
6H^2=\rho_c+\rho_m\,,
\end{eqnarray}
where 
\begin{eqnarray}
\label{rhoc}
\rho_c \equiv 
\frac{{\rm d}}{{\rm d}t}\left(\frac{\partial {\cal L}_c}
{\partial  \dot{N}}\right)+3H\frac{\partial {\cal L}_c}
{\partial  \dot{N}}-\frac{\partial {\cal L}_c}{\partial N}
-{\cal L}_{c} \biggr|_{N=1} \,.
\end{eqnarray}
$\rho_{m}$ is the energy density of the barotropic fluid, satisfying
\begin{eqnarray}
\label{rhom}
\dot{\rho}_{m}+3H(1+w_{m})\rho_{m}=0\,.
\end{eqnarray}
In what follows we shall consider the case with a fixed dilaton.
Then we have two dynamical equations (\ref{Hami})
and (\ref{rhom}) for our system.
We note that the variation of the action (\ref{action}) in terms of 
the scale factor $a$ gives rise to another equation, but this can be 
derived from Eqs.~(\ref{Hami}) and (\ref{rhom}) 
by taking a derivative with respect to $t$.
{}From Eq.~(\ref{rhoc}) we find that the energy density $\rho_c$
for type II \& heterotic strings is 
\begin{eqnarray}
\label{corre}
\rho_c &=& B [a_8H^8+a_c I^4+a_4H^4I^2+a_2H^2I^3+a_6H^6I
\nonumber \\
& & -J(a_5H^5+a_1HI^2+a_3H^3I)]\,,    
\end{eqnarray}
where $J=\ddot{H}+3H\dot{H}+H^3$.
One has $B=6c_3 \alpha'^3 e^{-6\phi}\zeta(3)$, 
$a_8 =-21, a_c=-3, a_4=-12, a_2=4, a_6=-24, a_5=-8,
a_1=-12, a_3=-24$ for type II string, and $B=6c_3 \alpha'^3 
e^{-6\phi}$, $a_8=-21\zeta(3)+35, a_c=-3\zeta(3)+15, a_4=-12\zeta(3)-6,
a_2= 4\zeta(3)+20, a_6=-24\zeta(3)+12, a_5=-8\zeta(3)+4,
a_1=-12\zeta(3)+60, a_3=-24\zeta(3)$ for heterotic string.
In the case of bosonic strings we have 
\begin{eqnarray}
\label{corre2}
\rho_c &=&
A (5H^6+2I^3-6HIJ) +B [(-21\zeta(3)+210)H^8
\nonumber \\
& &+(-3\zeta(3)+90)I^4-(12\zeta(3)+48)H^4I^2
\nonumber \\
&& +
(4\zeta(3)+120)H^2I^3 +(-24\zeta(3)+96)H^6I\nonumber \\
&&
+J \{(8\zeta(3)-32)H^5+(12\zeta(3)-360)HI^2
\nonumber \\
&&+
24\zeta(3)H^3I \}]\,.
\end{eqnarray}
where $A=24c_2\alpha'^2e^{-4\phi}$ and $B=6c_3 \alpha'^3 e^{-6\phi}$.

\section{The fate of dark energy universe}

In this section we study the cosmological evolution in dark energy 
universe for the above three classes of string curvature corrections.
Our main interest is the universe dominated by phantom dark energy, 
but we consider the case of cosmological constant as well.
We first study the effects of type II and heterotic corrections 
and then proceed to the bosonic correction.

\subsection{Type II and heterotic strings}

For the analysis of dynamics to follow, it would be convenient for us
to cast Eqs.~(\ref{Hami}) and (\ref{rhom}) 
with the correction term (\ref{corre}) in the form:
\begin{eqnarray}
\label{auto1}    
\dot{x} &=& y\,, \\
\dot{y} &=& [B\{a_8x^8+a_c I^4+a_4x^4I^2+a_2x^2I^3+a_6x^6I \nonumber \\
& & -(3xy+x^3)\xi \}+z-6x^2]/(B\xi)\,, \\ 
\label{auto3}    
\dot{z} &=& -3(1+w_m)xz\,,
\end{eqnarray}
where $x=H$, $y=\dot{H}$, $z=\rho_m$, $I=x^2+y$ and 
$\xi=a_5x^5+a_1xI^2+a_3x^3I$.
Here we shall consider the case with $w_{m } \ne -1$.
By setting $\dot{x}=0$, $\dot{y}=0$ and $\dot{z}=0$, we 
find the following de-Sitter fixed point:
\begin{eqnarray}
x_c=\left(\frac{6}{BD}\right)^{1/6}\,,~~
y_c=0\,,~~z_c=0\,,
\end{eqnarray}
where $D=a_c+a_2+a_{4}+a_{6}+a_{8}-a_{1}
-a_{3}-a_{5}$.
Since $D<0$ for both type II and heterotic strings, 
we do not have de-Sitter fixed points.

\begin{figure}
\includegraphics[height=2.8in,width=2.8in]{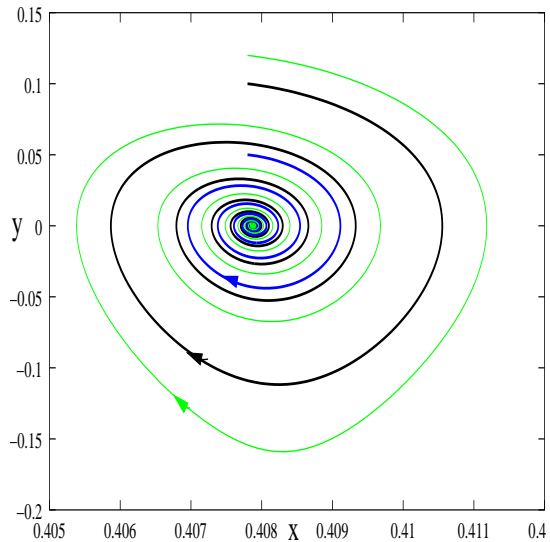}
\caption{\label{fig1}
The phase portrait for heterotic string in the case 
of $\rho_m\equiv \Lambda=1$ for several different 
initial conditions. The stable fixed point 
$x_c \equiv H_c=0.408$ corresponds to a 
de-Sitter solution which is a stable spiral.}
\end{figure}

When a cosmological constant $\Lambda$ is present
instead of $\rho_m$, corresponding to the equation of 
state $w_m=-1$, we find from Eq.~(\ref{Hami})
that there exists one de-Sitter solution which satisfies 
$\Lambda=6H^2-BDH^8$.
One can study the stability of this solution
by considering small perturbations $\delta x$ and 
$\delta y$ about the fixed point. We evaluate 
two eigenvalues for the matrix of perturbations
using the method in Ref.~\cite{CLW}.
For the type II correction we find that one eigenvalue is positive 
while another is negative, thereby indicating that the 
de-Sitter solution is not stable.
Meanwhile in the heterotic case the de-Sitter solution is either
a stable spiral (for smaller values of $\Lambda$, see Fig.~\ref{fig1}) 
or a stable node (for larger values of $\Lambda$).
We note, however, that this stable solution disappears
for the equation of state with $w_m \ne -1$.

\begin{figure}
\includegraphics[height=3.1in,width=3.1in]{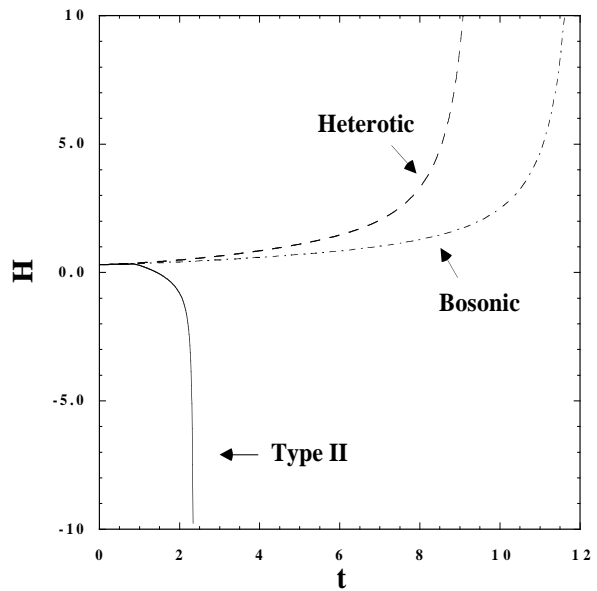}
\caption{\label{fig2}
The evolution of the Hubble rate $H$ in the presence of 
string curvature corrections and a phantom fluid with 
an equation of state: $w_m=-1.5$.
For the type II correction, 
the solution approaches $H=-\infty$ by crossing $H=0$, 
whereas in the heterotic and bosonic cases 
the Hubble rate grows toward infinity.
} 
\end{figure}

In order to understand the fate of the universe which is 
dominated by a phantom-type fluid, 
we solve autonomous equations 
(\ref{auto1})-(\ref{auto3}) numerically.
Figure \ref{fig2} shows the evolution of the Hubble rate
for a phantom-type fluid with $w_m=-1.5$
in the presence of string curvature corrections.
We find that in the type II case 
the Hubble rate begins to decrease because of the presence
of string corrections and it eventually diverges toward 
$-\infty$ after crossing $H=0$.
Thus the fate of the universe is characterized by 
a Big Crunch rather than a Big Rip.
Meanwhile in the heterotic case the Hubble rate 
continues to increase and diverges after a finite interval of time,
see Fig.~\ref{fig2}.
Thus the Big Rip singularity is inevitable
even when the heterotic string correction is present.

We also studied cosmological evolution for several different 
values of $w_m$ and for different initial conditions of
the Hubble rate.
For the type II correction we find that the solutions approach
the Big Crunch singularity for $w_m \lesssim -1.2$, whereas
they tend to approach the Big Rip 
singularity for $-1.2 \lesssim w_{m}<-1$.
Thus the fate of the universe depends upon the equation of
state for phantom dark energy.
For the heterotic case we find that the solutions reach to 
the Big Rip singularity independent of the values of 
$w_{m}~(<-1)$ and initial conditions of $H$.

\subsection{Bosonic string}

As demonstrated above, the type II and the heterotic string models
do not exhibit de-Sitter solutions for $w_m \ne -1$. However,
in the case of bosonic string, there exists a de-Sitter fixed point
which satisfies the relation
\begin{eqnarray}
\label{BodeSi}
B \left[76-12\zeta(3)\right]H^6+AH^4-6=0\,.
\end{eqnarray}
For example one has $H=0.709$ for $A=1/2$ and $B=3/4$.
By considering small perturbations around this solution and 
evaluating three eigenvalues of the $3 \times 3$ matrix for the system 
given by Eqs.~(\ref{auto1})-(\ref{auto3}) , 
we find that two of the eigenvalues are positive for 
$w_m<-1$ and one of them is positive for $w_m>-1$.
Therefore the de-Sitter solution characterized 
by Eq.~(\ref{BodeSi})  does not correspond to 
a stable attractor for $w_m \ne -1$.

In Fig.~\ref{fig2} we plot the evolution of the Hubble rate 
with bosonic string corrections in the presence of a
phantom-type fluid with $w_m=-1.5$. 
The Hubble rate continues to grow
and diverges with a finite time as in the case of 
heterotic string. 
We also run our numerical code for different values of 
$w_m$ and find that the solutions approach the 
Big Rip singularity for $w_m<-1$.

\begin{figure}
\includegraphics[height=3.1in,width=3.1in]{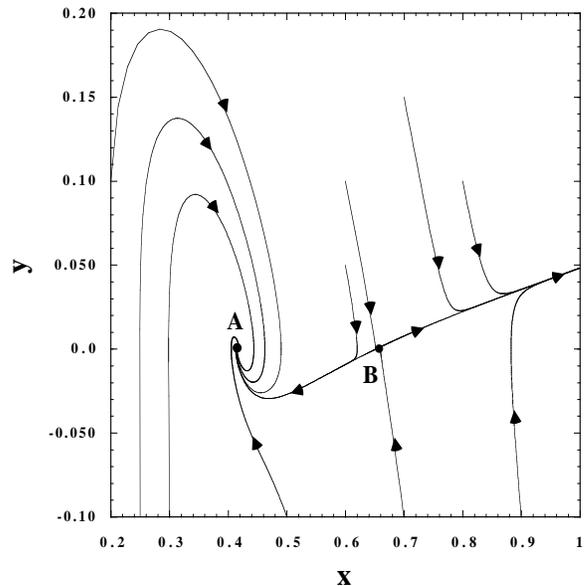}
\caption{\label{fig3}
The phase portrait for bosonic string in the presence  
of a cosmological constant ($\rho_m \equiv \Lambda=1$)
with $A=1/2$ and $B=3/4$.
There exist two de-Sitter fixed points.
The point A [$x_c=0.417$]  corresponds to 
a stable spiral, whereas the point B [$x_c=0.652$]
is a saddle.
}
\end{figure}

When a cosmological constant $\Lambda$ is present 
instead of $\rho_m$, de-Sitter solutions satisfy
\begin{eqnarray}
\Lambda=6H^2-AH^6-B\left[76-12\zeta(3)\right]H^8\,.
\end{eqnarray}
There exist two solutions for this equation provided that $\Lambda$
ranges in the region
$0<\Lambda<f(H_{M}) \equiv 6H_M^2-AH_M^6-B\left[76-12\zeta(3)\right]H_M^8$,
where $H_{M}$ is the solution for 
$3AH_{M}^4+4B[76-12\zeta(3)]H_{M}^6=6$.
For example $H_{M}=0.56$ and $f(H_M)=1.42$
for $A=1/2$ and $B=3/4$.
When $\Lambda=1$, one has two de-Sitter solutions
characterized by $H=0.417$ and $H=0.652$.
We evaluate two eigenvalues of the $2 \times 2$ matrix 
for perturbations $\delta x$ and $\delta y$ around the fixed points. 
We find a complex conjugate pair of eigenvalues with negative real part
for $H=0.417$ making the fixed point a stable spiral.
As for the second second critical point corresponding to $H=0.652$, 
one of the eigenvalues turns out to be 
positive, which means that 
the de-Sitter solution is unstable$-$ a saddle 
in this case (see Fig.~\ref{fig3}).

%
\section{Summary}

In this paper we studied the effect of higher-curvature corrections
in low-energy effective string theory on the cosmological dynamics
in the presence of dark energy fluid. Since the existence of a
phantom fluid leads to the growth of the Hubble rate, 
the energy scale of the universe may reach the Planck scale in future.  
This means that string curvature corrections can be very important
to determine the dynamical evolution of the universe.

We have considered string corrections up to quartic in curvatures
for three type of string theories-- (i) type II, (ii) heterotic, and
(iii) bosonic strings. In our analysis the contribution of the Gauss-Bonnet term 
does not affect the cosmological dynamics in $D=4$ dimensions,
since the dilaton and the modulus fields are fixed.
For the fluid with an equation of state characterized by 
$w_m \ne -1$, we find that 
de-Sitter solutions do not exist for type II and heterotic string 
corrections. There is a de-Sitter solution for the bosonic string
even for $w_m \ne -1$, but this is found to be unstable. 
When the equation of state for fluid is that of a cosmological 
constant ($w_m=-1$), we find that 
stable de-Sitter solutions exist for heterotic and bosonic strings. 

We ran our numerical code to study the effect of string corrections 
around the Big Rip when a phantom fluid is present.
In the type II case we found that the solutions approach 
the Big Crunch singularity ($H \to -\infty)$ after crossing $H=0$
when the equation of state for dark energy is $w_m \lesssim -1.2$.
Meanwhile the Hubble rate diverges toward $H \to +\infty$
with a finite time for heterotic and bosonic corrections, 
which implies that the Big Rip singularity is difficult to 
be avoided in these cases. 
The divergent behavior of the Hubble rate for $w_{m}<-1$ 
is associated with the fact 
that there are neither stable de-Sitter nor stable Minkowski 
attractors for the types of the corrections we considered.

In the present letter we restricted our attention to purely
geometrical effects assuming that non-perturbative potentials
may arise allowing to freeze dilaton and modulus fields. Compactifications
from higher dimensions to 4-dimensional space time result in residual
modulus fields which are related to the radii of internal space. In
general a modulus field is dynamical and interacts with higher-order
curvature terms. The same is true for the dilaton field which is related
to string coupling $g_s$. These
may give rise to non-trivial effects, for instance, even the GB curvature
invariant which is purely topological in 4 dimensions, 
does not vanish in the presence of dynamical  
modulus and dilaton fields. 
Such a scenario would have important implications for
future evolution of the dark energy universe. 
Very recently, cosmological dynamics based upon effective string theory 
action was investigated with dynamically evolving modulus and dilaton fields
in the presence of second-order curvature corrections \cite{Gian} 
(see Ref.~\cite{Odintsov} on the related theme). 
It was demonstrated that the second-order curvature correction to Einstein-Hilbert 
action can significantly modify the structure of future singularities in
dark energy universe. It is therefore important, though technically cumbersome,
to extend the analysis of the present letter to the case of dynamical modulus/dilaton
fields.
                                                                                
A comment is also in order about the distinct features that different string
models exhibit in cosmological dynamics. With fixed modulus/dilaton,
the evolution of phantom dark energy universe is clearly distinguished 
depending upon string models, namely, the fate of such a universe is 
Big Crunch for the type II superstring whereas it is a Big Rip for 
bosonic and heterotic strings.
This distinction mainly comes from the difference of the  
coefficients in Eqs.~(\ref{corre}) and (\ref{corre2}).
In the type II case $\rho_{c}$ becomes negative,
which counteracts the energy density of the phantom fluid.
This property is crucially important to avoid the Big Rip singularity
as pointed out in Ref.~\cite{NOT}. 
We note that this behavior also appears in the presence of a 
dynamical modulus field with second-order 
string corrections \cite{Gian}.
It is really of interest to investigate how the final fate of the 
universe is changed when dynamical modulus/dilaton 
fields couples to third/fourth-order string
curvature corrections.
We hope to address this issue in future work.

In addition, the higher-order curvature contributions used in our 
description have inbuilt ambiguities related to 
particular metric redefinitions. It would be important to investigate whether 
or not these ambiguities can
lead to different fate of cosmological evolution.

\section*{ACKNOWLEDGEMENTS}
We thank O.~Bertolami, G.~Calcagni, F.~Fattoev, T.~Naskar, S.~Nojiri 
and T.~Padmanabhan for useful discussions. 
A.T. acknowledges support from IUCAA's ``Program for
enhanced interaction with the Africa$-$Asia$-$Pacific Region''.
The work of S.T. was supported by JSPS (No.\,30318802).

\section*{APPENDIX: Calculation of curvature tensors}

For the metric (\ref{metric}) the non-zero components of Christoffel 
symbols are
\begin{eqnarray}
\Gamma^0_{\mu \mu}=\frac{a\dot{a}}{N^2}\,,~~~
\Gamma^{\mu}_{0 \mu}=\Gamma^{\mu}_{\mu 0}
=\frac{\dot{a}}{a}\,,~~~
\Gamma^0_{00}=\frac{\dot{N}}{N}\,, \nonumber 
\end{eqnarray}
where $\mu=1,2,\ldots d$.
By using the formula
\begin{eqnarray}
R^L_{SMN}=\partial_M\Gamma^L_{SN} - \partial_N\Gamma^L_{SM} +
\Gamma^L_{MR}\Gamma^R_{NS}
- \Gamma^L_{NR}\Gamma^R_{MS}\,, \nonumber 
\end{eqnarray}
we find that the non-zero components of Riemann tensors are
\bea
\label{cur}
& &R^0_{\mu 0\mu}=-R^0_{\mu \mu 0}=\frac{a\ddot a}{N^2}-
\frac{a\dot{a}\dot{N}}{N^3}\,, \nonumber \\
& &R^\mu_{00\mu}=-R^\mu_{0 \mu 0}=\frac{\ddot a}{a}-\frac{\dot{a}}{a}
\frac{\dot{N}}{N}\,,~~R^\mu_{\nu\mu\nu}=-R^\mu_{\nu\nu\mu}=\frac{\dot a^2}{N^2}\,.
\nonumber
\eea
These give
\begin{eqnarray}
& &R^{\mu 0}_{0\mu}=R^{0 \mu}_{\mu 0}
=-R^{\mu 0}_{\mu 0}=-R^{0 \mu}_{0 \mu}=
\frac{1}{N^2}\left(
\frac{\ddot a}{a}-\frac{\dot{a}}{a}\frac{\dot{N}}{N}\right)\,, \nonumber \\
& &R^{\mu\nu}_{\nu \mu}=-R^{\mu\nu}_{\mu \nu}=
\frac{1}{N^2}\frac{\dot a^2}{a^2}\,. \nonumber
\end{eqnarray}
Noting the relation $R_{MN}=R^L_{MLN}$, we obtain
\begin{eqnarray}
R_{00}=-d \left(\frac{\ddot{a}}{a}-\frac{\dot{a}}{a}
\frac{\dot{N}}{N}\right)\,,~
R_{\mu \mu}=\frac{a\ddot{a}}{N^2}-\frac{a\dot{a}\dot{N}}{N^3}
+(d-1)\frac{\dot{a}^2}{N^2}\,. \nonumber
\end{eqnarray}
We also find 
\begin{eqnarray}
& & R_{0 \mu \mu 0}=R_{\mu 0 0 \mu}=
-R_{0 \mu 0 \mu}=-R_{\mu 0 \mu 0}=
a\ddot{a}-a\dot{a}\frac{\dot{N}}{N}\,, \nonumber \\
& & R_{\mu \nu \mu \nu}=-R_{\mu \nu \nu \mu}=
\frac{a^2\dot{a}^2}{N^2}\,, \nonumber \\
& & R^{0 \mu \mu 0}=R^{\mu 0 0 \mu}=
-R^{0 \mu 0 \mu}=-R^{\mu 0 \mu 0}=
\frac{1}{N^4a^3}\left(\ddot{a}
-\dot{a}\frac{\dot{N}}{N}\right), \nonumber \\
& & 
R^{\mu \nu \mu \nu}=-R^{\mu \nu \nu \mu}=
\frac{\dot{a}^2}{N^2a^6}\,. \nonumber
\end{eqnarray}
These relations are used to evaluate the correction term ${\cal L}_{c}$.

We should note that the corrections can be easily computed
in case of de-Sitter symmetry [$R_{abcd}=\Lambda \left( g_{ac}g_{bd}
-g_{ad} g_{bc} \right)$], as
\begin{eqnarray}
 \label{dL1}
{\cal L}_3 &=& 4\Lambda^3 d(d+1) \,, \\
{\cal L}_{41}&=& -3\zeta(3) \Lambda^4 (d-1)d(d+1) \,, \\
{\cal L}_{42}&=& \frac{1}{2}\Lambda^4 (d-7)d^2(d+1) \,, \\
\label{dL3}
{\cal L}_{43}&=& -4\Lambda^4 d(d+1)\left[1-d(d-9)\right]\,.
\end{eqnarray}
These coincide with Eqs.~(\ref{L3})-(\ref{L43}) by setting $N=1$ and 
$H^2=\Lambda$.


\end{document}